# Giant and Broadband THz and IR Emission in Drift-biased Graphene-Based Hyperbolic Nanostructures

L. Wang[1], N. K. Paul[1], J. Hihath[2], and J. S. Gomez-Diaz[1]*

[1] Department of Electrical and Computer Engineering, University of California Davis

[2] Biodesign Center for Bioelectronics and Biosensors, School of Electrical, Computer and Energy Engineering, Arizona State University

ABSTRACT. We demonstrate that Cherenkov radiation can be manipulated in terms of operation frequency, bandwidth, and efficiency by simultaneously controlling the properties of drifting electrons and the photonic states supported by their surrounding media. We analytically show that the radiation rate strongly depends on the momentum of the excited photonic state, in terms of magnitude, frequency dispersion, and its variation versus the properties of the drifting carriers. This approach is applied to design and realize miniaturized, broadband, tunable, and efficient terahertz and far-infrared sources by manipulating and boosting the coupling between drifting electrons and engineered hyperbolic modes in graphene-based nanostructures. The broadband, dispersive, and confined nature of hyperbolic modes relax momentum matching issues, avoid using electron beams, and drastically enhance the radiation rate – allowing that over 90% of drifting electrons emit photons. Our findings open a new paradigm for the development of solid-state terahertz and infrared sources.

The generation and control of electromagnetic waves has helped shape the development of our society. Approaches like thermal radiation (light bulbs), thermionic emission (vacuum tubes), stimulated emission (lasers) and electroluminescence (LEDs) have been incorporated into everyday technologies covering the realms of electronics and photonics. Between these two regimes lies the terahertz (THz) and far-infrared (IR) band (∼ 0.5 to 20 THz), which offer unique opportunities to enable communication systems with terabit per second data rates, ultra-fast and miniaturized processing systems, high-resolution imaging for inspection and screening, and innumerable biomedical and sensing devices. Unfortunately, our society is not fully enjoying the benefits of these and other applications because the limited response of materials in this frequency band prevents efficient generation of THz and far-IR light in a simple and portable way [1, 2]. Indeed, state-of-the-art sources suffer from shortcomings in terms of bandwidth, power, size, and complexity which may require operating at low temperature [3-5] or the presence of optical pumps [6] and large magnets [7].

Another fundamental mechanism for generating light, Cherenkov radiation [8-10], has been limited to niche applications such as medical imaging, dosimetry, and particle detection because it typically requires highly energetic electrons produced by nuclear reactors, particle accelerators, or radionuclide generators [11]. Recently, light-matter interactions with photonic quasiparticles [12] and the possibility to obtain Cherenkov radiation in advanced photonic systems has been explored. For instance, it has been experimentally demonstrated that low-energy electron beams propagating near hyperbolic metamaterials (HMTMs) [13-14] are able to generate radiation in the visible [15]. HMTMs allow control over the electron velocity threshold, intensity, and frequency of the emission by manipulating their anisotropy and local density of states. Other works have also studied this possibility using different plasmonic structures, such as thin metallic films [16] or graphene [17,18]. The main challenge of these devices is the need of external electron beams travelling near the nanostructures, which is difficult to achieve in an integrated fashion. Additionally, it has been predicted that drifting electrons in graphene can couple to surface plasmon polaritons [19] whereas several works have highlighted the coupling between electrons and plasma waves in 2D electron gasses found in transistors and their potential applications [20-23]. Such processes are somewhat inefficient and narrowband, and thus it is challenging to apply them to construct sources in practice. Even more recently, drifting electrons in graphene have been shown to enable a plasmonic Doppler effect associated with strong electromagnetic nonreciprocity [24-27]. In all these platforms, Cherenkov radiation was an afterthought in the sense that it was studied once the electromagnetic properties of the system were fixed, and therefore it was not possible to manipulate the properties of the resulting radiation.

In this letter, we demonstrate that Cherenkov emission can be enforced at desired wavelengths and drastically boosted in terms of efficiency and bandwidth by tailoring at the nanoscale the photonic density of states of the media that surrounds drifting electrons (Fig. 1a). By imposing energy and momentum conservation, we develop an approach to determine the electromagnetic properties of photonic states that allow their excitation from low-energy drifting electrons flowing *through* the media. Our analytical study reveals that the radiation rate strongly depends on the momentum of the excited photonic states, in terms of (i) magnitude and frequency dispersion; and (ii) how it varies with the properties of drifting electrons. These modes can be implemented in practice using a wide variety of photonic configurations, including cavities, resonators, or nanostructures [28]. To illustrate this broad concept, we propose to harness emission at long wavelengths by exploiting the coupling between drifting electrons and hyperbolic photonic modes in nanostructures made of 2D materials. Graphene [29] is an ideal candidate for this application: it is a highly conductive material whose plasmonic nature at THz/IR frequencies allows hyperbolic modes to be engineered in this range when the material is stacked with dielectrics [30-33] or patterned at the nanoscale [34-37]. Electrons flowing within graphene can be energy- and momentum-matched with broadband yet confined hyperbolic modes to enable a high-rate of photon emission through Cherenkov radiation. Compared to electron-photon coupling in standard plasmonic structures, such as pristine graphene [19], the use of hyperbolic nanostructures boosts the spontaneous radiation rate over three orders of magnitude and drastically enhances the emission bandwidth to operate from terahertz to mid-IR frequencies. Our theoretical formalism predicts that over 90% of drifting electrons will emit photons, thus opening new and excited venues for the development of miniaturized, broadband, and efficient sources

Let us consider a graphene sheet longitudinally biased with a DC voltage $V_{DC}$ that induces low-energy drifting electrons with velocity $\boldsymbol{v}_d = v_d \hat{y}$ and energy $E_i = \hbar v_F |\mathbf{k}_i|$ (Fig. 1a), where $\hbar$ and $v_F \approx 10^6 \text{ms}^{-1}$ are the reduced Planck constant and Fermi velocity, respectively. The goal is to engineer the electromagnetic properties of the media surrounding graphene in terms effective permittivity ($\bar{\bar{\varepsilon}}$) and permeability ($\bar{\bar{\mu}}$) to enforce Cherenkov radiation at a frequency $\omega$ (Fig. 1b). This process is illustrated in Fig. 1c using an energy-momentum diagram: drifting electrons spontaneously relax into states with energy $E_f = \pm \hbar v_F |\mathbf{k}_f|$ and emit photons with momentum $\mathbf{q}$ and energy $\hbar\omega$. The electronic states enabling this process exhibit a momenta with magnitude $|\mathbf{k}_f| = \frac{|E_i - \hbar\omega|}{\hbar v_F}$ that displays a circular shape in the momentum space of Fig. 1c (dashed red line in Fig. 1c) due to the linear dispersion of graphene. Additionally, graphene's Fermi level $\mu_c$ must be lower than the energy of the relaxed states $E_f$ to keep these states vacant. To enable the desired photon emission, the media surrounding drift-biased graphene should support a photonic state $\mathbf{q}$ that fulfills momentum conservation, i.e., $\hbar \mathbf{k}_i = \hbar \mathbf{k}_f + \hbar \mathbf{q}$. Fig. 1d illustrates this process using an isofrequency (IFC) contour in the photonic momentum space [28]. In this diagram, the photonic states that enable Cherenkov emission yield a circle centered at $\left(0, \frac{|\mathbf{k}_i|}{q_0}\right)$ and with radius $\frac{|\mathbf{k}_f|}{q_0}$, where $q_0 = \omega/c$ is the free space wavenumber at $\omega$. We denote the momenta of such states as "electron-photon coupling space". Any media that supports a state within this space (as the green dot in Fig. 1d) would allow spontaneous coupling between drifting electrons and photons through Cherenkov emission. The realm of metamaterials provides large flexibility to engineer artificial media at targeted wavelengths with desired photonic states in the form of bulk waves or even surface plasmon polaritons. Considering a linear, lossless, local, and anisotropic graphene-based metasurface and applying the Fermi's golden rule [38] as described in the Supplementary Information, it is possible to analytically calculate the radiation rate into a specific photonic state $\mathbf{q}_s$ as

$$\Gamma_{\mathbf{q}_s} = \frac{e^2 v_F}{\pi \hbar^2 \omega^2} |\mathbf{E}_\parallel|^2 S \frac{dq_t}{d|\mathbf{k}_f|} \frac{dq_p}{d\omega} \sin^2\left(\theta - \frac{\pi}{4} - \frac{\varphi}{2}\right), \qquad (1)$$

where $\alpha$ is the fine structure constant, $\omega$ is the photon frequency, $\varepsilon_0$ is the free space permittivity, c is the speed of light, $\theta = \arctan\left(\frac{E_y}{E_x}\right)$, $\varphi = \arctan\left(\frac{k_y}{k_x}\right)$, $\mathbf{q}_s = q_t \hat{t} + q_p \hat{p}$ with $\hat{t}$ and $\hat{p}$ being unit vectors in the direction tangential and perpendicular to the dispersion relation $\mathbf{q}(\omega)$ of the structure evaluated at $\mathbf{q}_s$, $\mathbf{E}_\parallel$ is the in-plane electric field distribution of the photonic state [39], and $S \cdot |\mathbf{E}_\parallel^2|$ is related to the in-plane energy of the excited mode. This formalism is valid for interband transitions – i.e., drifting carriers relaxing from the conduction to the valence band – and has been expanded in the Supplementary Information [40] to consider intraband transitions and the presence of loss. Eq. (1) reveals that the emission rate depends on the properties of the excited photonic state in terms of energy and momentum. Specifically, it is directly proportional to (i) the frequency dispersion of the supported state; and (ii) the variation of the momentum of the excited mode versus the momentum of the relaxed electron – i.e., how the excited state changes as the properties of the drifting carriers evolve. Therefore, it is possible to enforce and manipulate Cherenkov radiation in terms of operation frequency, bandwidth, and efficiency by simultaneously controlling the properties of drifting electrons and the photonic states supported by the surrounding media.

To realize broadband and efficient THz and IR emitters, we propose the unique combination of drifting electrons flowing through graphene and the hyperbolic dispersion of tailored nanostructures. Graphene-based hyperbolic metasurfaces (HMTSs) are excellent candidates for implementing this type of sources as they exhibit a high density of confined photonic states over a broad frequency range in the THz/IR band and can be fabricated by nanostructuring a graphene sheet [34-37]. The design process to construct this type of plasmonic emitters is as follows. First, we consider that electrons with an initial energy $E_i$ are flowing through graphene. In this example, we set a conservative energy of $E_i = 0.03$eV ($v_d \approx 0.1\ v_F$) – note that drift velocities up to $v_d \approx 0.9\ v_F$ have been experimentally reported in graphene encapsulated in hBN [41]. Second, we calculate the electron-photon coupling space versus frequency in the photonic momentum diagram (grey line in Fig. 2a). That space represents the potential photonic states that can be excited in this platform. Other states at desired wavelengths can be obtained by manipulating the initial energy of the flowing electrons. And third, we design a hyperbolic metasurface able to support photonic states that intersect with the electron-photon coupling space. Here, we consider an array of densely packed graphene strips with width $W = 50$nm and periodicity $L = 100$nm. This structure can be analyzed using effective medium theory [35] and supports confined, hyperbolic modes over a large bandwidth in the IR (blue line in Fig. 2a). Photonic states excited by drifting electrons are determined by the intersection between the modes supported by the nanostructure and the electron-photon coupling space, as illustrated by magenta lines in Fig. 2a. The broadband, dispersive, and confined nature of hyperbolic modes permits the generation of photons throughout the IR band ($\sim 11-30$ THz). This is in stark contrast to a common graphene sheet [19] in which drifting electrons are unable to efficiently excite broadband photonics modes, as shown in Fig. 2b. Fig. 2c plots the IFC of the hyperbolic metasurface and graphene sheet together with the electron-photon coupling space. Results show the evolution of the electron-photon coupling space versus frequency. Interestingly, hyperbolic metasurfaces may simultaneously support forward (central panel) and even backward emission (right panel) depending on the momentum of the excited states (magenta dots). Even though emission also appears in a pristine graphene layer, there it is very narrowband and inefficient, thus challenging to be observed in practice. In the Supplementary Information, we further study THz/IR generation in various HMTSs for a wide range of initial energies of the drifting electrons. Our study confirms that the emission is robust, broadband, and can be manipulated with the geometrical parameters that define the nanostructures.

This spontaneous emission process can also be understood using energy-momentum diagrams, as illustrated in Fig. 3 for a HMTS (a) and a graphene sheet (b). Specifically, energy and momentum conservation determine the electronic states in which drifting electrons (blue dot) can relax to. In case of HMTSs, the electron-photon coupling space translates into a relatively large set of available electronic states that would enable the emission process (solid red line). As a result, a drifting electron has many states to relax to, thus leading to a broadband emission. Note that the distribution of these states as well as the properties of the associated radiation significantly changes with the photonic states supported by the surrounding media [40]. This situation is quite different in case of a graphene layer because drifting electrons can only relax into a very narrow set of electronic states. As a result, the emission process is narrowband and severely hindered.

Fig. 4 shows the total radiation rate and spectrum distribution of the emission predicted in a drift-biased graphene-based HMTS (solid line) and in a pristine graphene layer (dashed line). Results are computed for a wide range of hyperbolic metasurfaces, accounting for applied gate voltage (i.e., graphene's chemical potential), velocity of drifting carriers, presence of loss (modelled through graphene's relaxation time [40]), and geometrical dimensions of the strip array (i.e., width and periodicity). It should be noted that the total radiation at a given frequency may be distributed into several photonic states that propagate toward different directions (Fig. 2c) [40]. In all cases, the presence of hyperbolic modes drastically boosts the overall efficiency of the emission process compared to the case of pristine graphene [19] while simultaneously enhancing the bandwidth from the mid-IR to the THz region. Such response can easily be understood from Eq. (1): hyperbolic modes are more confined than common surface plasmons in graphene and are inherently dispersive, thus leading to a much larger radiation bandwidth. On one end, the lower cut-off of the emission is directly proportional to the velocity of the drifting electrons ($f_c \propto v_d$ [40]) and thus can easily be manipulated in practice. Fig. 4a shows that the radiation onset can be tailored to appear at ~3 THz by decreasing $v_d$. It should also be noted that the rate is enhanced close to the emission cut-off frequency, which is attributed to a rapid variation of the coupling between the excited photonic state and drifting electrons [i.e., $\frac{dq_t}{d|\mathbf{k}_f|}$ term in Eq. (1)]. On the other, the higher cut-off frequency of the emission is determined by the metasurface topological transition, which in this type of HMTS can be controlled through the geometrical W/L ratio. In case that such topological transition appears at the near-IR band or even shorter wavelengths, phenomena such as interband transitions in graphene's conductivity [29] or the metasurface nonlocality [41] may dominate and determine the radiation cut-off. Our results confirm that the emission rate is resilient against the presence of loss and robust versus variations in the energy of the drifting electrons as well as graphene's quality and doping. The metasurface dispersion relation can be tailored through geometrical parameters (i.e., W/L ratio) to maximize its frequency dispersion, which in turns enhances further the radiation rate (Fig. 4d). Optimized geometries provide broadband emission rates over three orders of magnitude compared to the ones found in pristine graphene, an emission level that can be observed in experimental set-ups. A simple two-state system model [9, 38] permits to characterize this spontaneous emission process and approximately calculate how many drifting electrons will effectively emit photons. Considering a realistic $3\mu m$ long HMTS with made of graphene strips ($W/L = 0.5$) with relaxation time $\tau = 0.1$ps that is biased to support drifting electrons with a velocity $v_d = 0.1 v_F$, our approach predicts an outstanding efficiency over 90% which is drastically reduced to ~0.6% in the case of unpatterned graphene. Even though this model is approximate, as it neglects the excitation of phonons and other scattering mechanisms, it highlights the tremendous potential of the proposed platform to develop solid state THz and IR sources.

In conclusion, we have proposed a new approach to harness Cherenkov radiation at desired wavelengths by manipulating the photonic states supported by the media surrounding drifting electrons. We have developed an analytical formalism to model this phenomenon in anisotropic metasurfaces that unveils the relation among radiation efficiency, drifting electrons, and the momentum and frequency dispersion of the supported modes. Rooted on this framework, we have proposed drift-biased graphene-based hyperbolic metasurfaces as a promising candidate to implement plasmonic sources that are (i) broadband, covering from the THz up to the mid-IR frequency range; (ii) highly efficient; (iii) tunable, because the radiation onset is determined by the applied DC voltage; and (iv) miniaturized and easy to realize in practice, as they do not require the presence of optical pumps beams or magnetic fields. Moving beyond, we envision that this concept can be applied to develop a wide variety of solid-state THz and IR sources generating either surface plasmon polaritons or plane waves in free-space. For instance, instead of hyperbolic metasurfaces, such sources can rely on bulk graphene-based hyperbolic metamaterials [30-33] in which drifting electrons couple to bulk hyperbolic modes that will be subsequently radiated to free-space in the form of plane waves using mechanisms such as diffraction gratings or nanoparticles located at the outer layer. Additionally, configurations such as resonators and cavities can provide the photonic states required to enforce desired radiation whereas graphene can be substituted by thin layers of gold/silver [43-44] or other 2D conductive materials [45-46], thus opening exciting opportunities to construct efficient solid-state THz and IR sources.

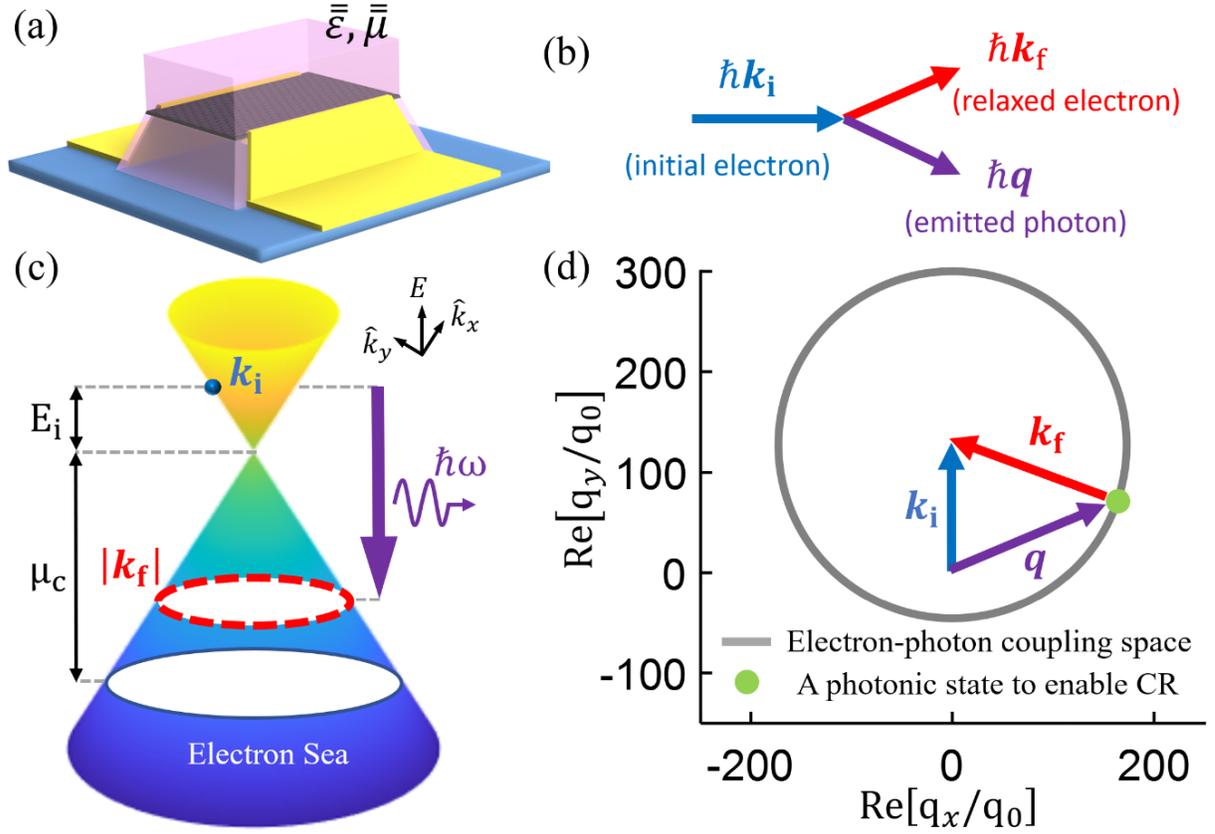

**Figure 1**. Tailoring Cherenkov radiation in drift-biased graphene-based nanophotonic devices. (a) Schematic of the structure. Graphene is embedded within an artificial media with effective permittivity and permeability $\bar{\bar{\varepsilon}}$ and $\bar{\bar{\mu}}$, respectively. (b) Schematic of the emission process. (c) Energy-momentum diagram. A drifting electron in graphene with energy $E_i$ and momentum $\boldsymbol{k_i} = k_y \hat{y}$ spontaneously relaxes into a state with energy $E_f$ and momentum $|k_f|$ while emitting a photon with energy $\hbar\omega$ and momentum $\mathbf{q}$. (d) Photonic states of the artificial media that would enable emission at $\omega$ (gray line). Any media supporting a state $\mathbf{q}$ within this electron-photon coupling space will enable Cherenkov emission. $q_0 = \omega/c$ is the free space wavenumber at $\omega$.

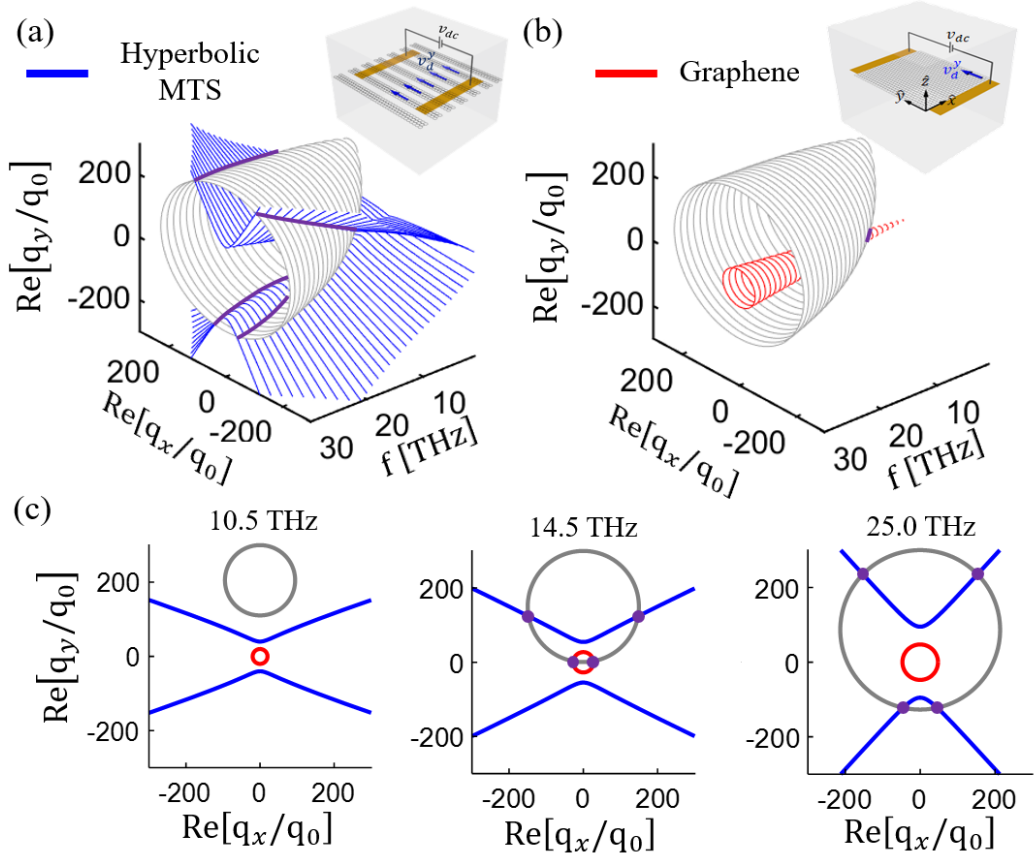

**Figure 2**. Photonic momentum diagrams to describe Cherenkov emission in a drift-biased graphene-based hyperbolic metasurface (a) and in a drift-biased graphene layer (b). The panels show the dispersion relation of the devices together with the electron-photon coupling space (grey lines) that would enable radiation. Intersection between supported modes and the electron-photon coupling space (purple lines) indicates the excited photonic states. (c) IFC of the top panels at different frequencies. Graphene's chemical potential is set to $\mu_c = -0.15$eV and the initial energy of drifting electrons is $E_i = 0.03$eV ($v_d \approx 0.1 v_F$). The lossless metasurface is composed of an array of packed graphene strips with width $W = 50$nm and periodicity $L = 100$nm.

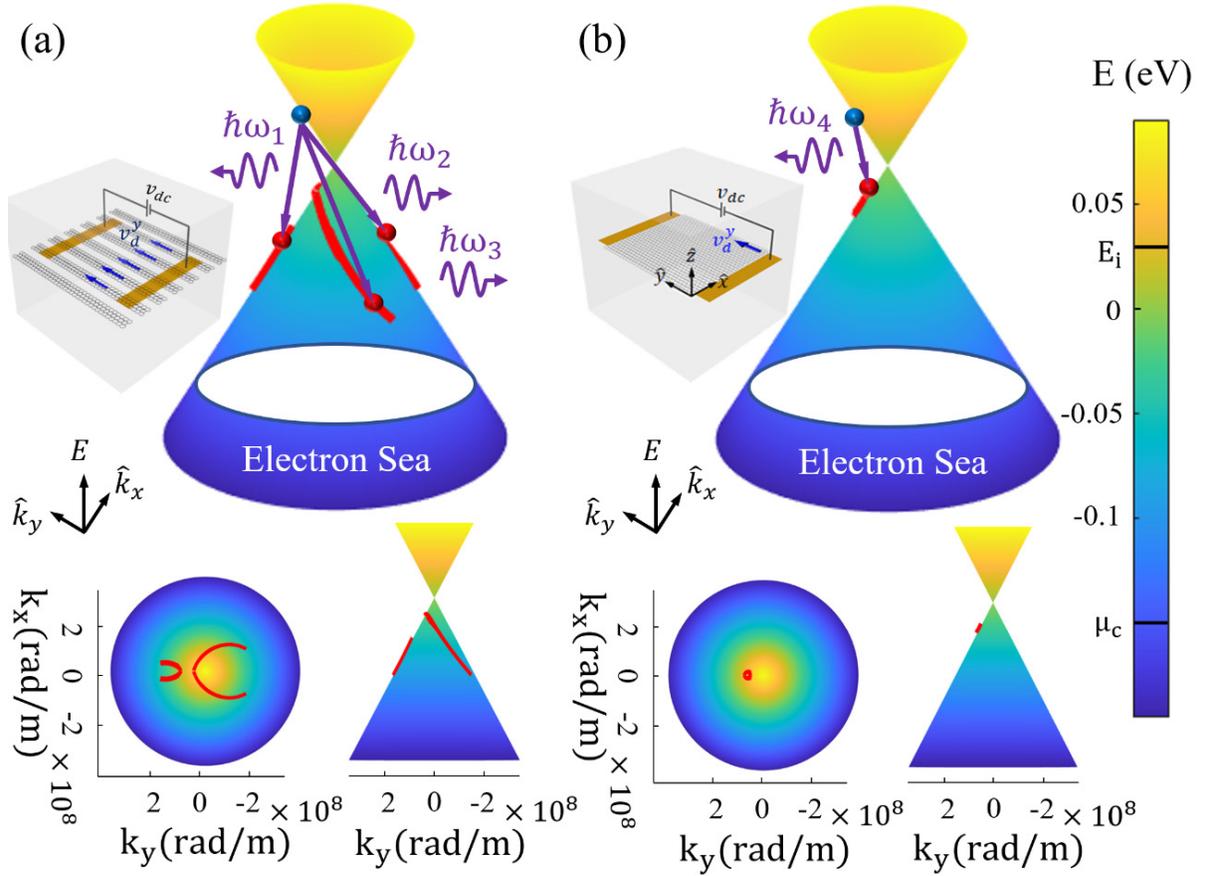

**Figure 3**. Electronic energy-momentum diagrams to describe Cherenkov emission in a drift-biased hyperbolic metasurface (a) and in a drift-biased graphene layer (b). Red lines show the allowed states in which drifting electrons (blue ball) can relax into while emitting a photon. Magenta arrows illustrate the emission process for specific cases. Other parameters are as in Fig. 2.

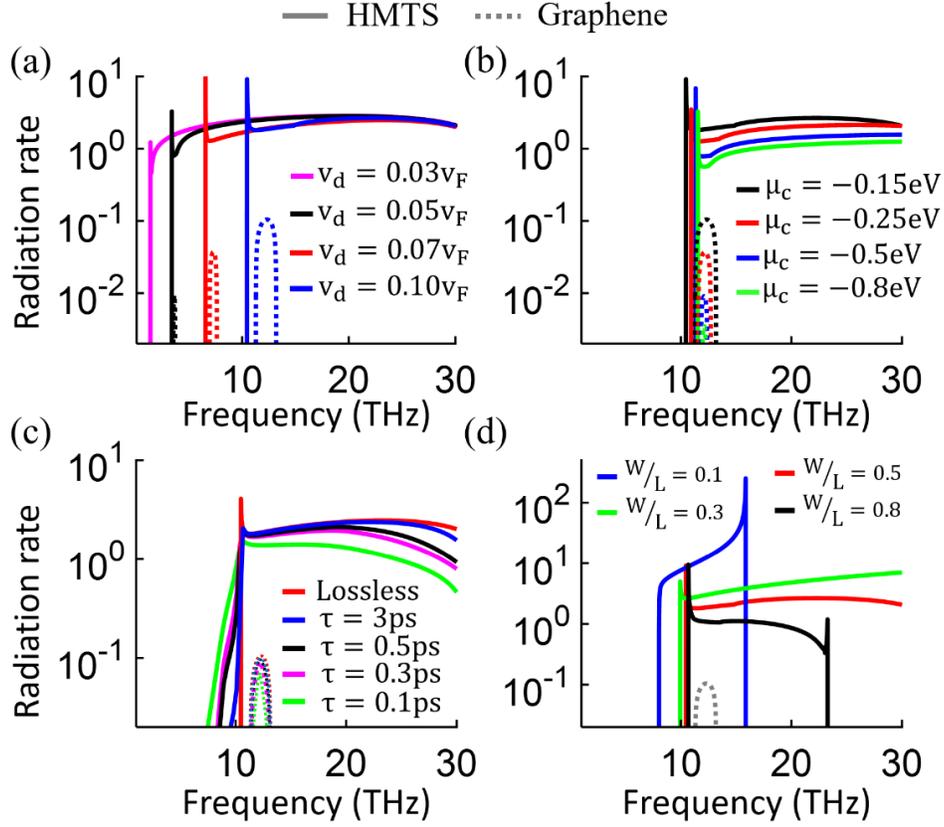

**Figure 4**. Radiation rate and spectrum distribution of the Cherenkov emission predicted in the drift-biased hyperbolic metasurface (solid lines) and in the drift-biased graphene layer (dashed lines) described in Fig. 2. Results are plotted versus frequency for different values of the carriers' velocity (a) and properties of the hyperbolic metasurface, including (b) graphene's chemical potential controlled through a gate bias; (c) loss, expressed through graphene's relaxation time $\tau$ [40]; and (d) geometrical dimension of the hyperbolic structure in terms of strip width W over periodicity, L [40]. Values for non-swept parameters are as in Fig. 2.

**Supporting Information**

Provides a detailed description of the proposed quantum formalism, including the development of the radiation rate formula shown in Eq. (1), as well as additional studies regarding the cut-off frequency of the emission and the influence of parameters as graphene's relaxation time and chemical potential, applied bias, geometrical dimensions, and electron velocity on the emission process.

**Corresponding Author**

J. S. Gomez-Diaz. E-mail: jsgomez@ucdavis.edu

**Acknowledgements**

This work was supported by the Keck Foundation and by the National Science Foundation with a CAREER Grant No. ECCS-1749177.

**References**

[1] P. H. Siegel, IEEE Trans. Microw. Theory Tech. **50**, 3, 910–928 (2002).

[2] M. Tonouchi, Nat. Photonics **1**, 2, (2007).

[3] J. Faist, F. Capasso, D. L. Sivco, C. Sirtori, A. L. Hutchinson, and A. Y. Cho, Science **264**, 553-556 (1994).

[4] M. A. Belkin and F. Capasso, Phys. Scr. **90**, 118002 (2015)

[5] K. Vijayaraghavan, Y. Jiang, M. Jang, A. Jiang, K. Choutagunta, A. Vizbaras, F. Demmerle, G. Boehm, M. C. Amann, M. A, Belkin, Nat. Commun. **4**, 2021 (2013).

[6] R. W. Boyd, *Nonlinear Optics*, 3rd ed. (Academic Press, 2008).

[7] J. H. Booske, R. J. Dobbs, C. D. Joye, C. L. Kory, G. R. Neil, G.-S. Park, J. Park, and R. J. Temkin, IEEE Trans. Terahertz Sci. Technol. **1**, 1 (2011).

[8] V. L. Ginzburg, Zh. Eksp. Teor. Fiz. **10**, 589 (1940)

[9] D. Jackson, *Classical Electrodynamics*, 3rd ed. (Wiley, 1998).

[10] I. Kaminer, M. Mutzafi, A. Levy, G. Harari, H. H. Sheinfux, S. Skirlo, J. Nemirovsky, J. D. Joannopoulos, M. Segev, and M. Soljacic, Physical Review X **6**, 011005 (2016).

[11] T. M. Shaffer, E. C. Pratt, and J. Grimm, Nat. Nanotechnol. **12**, 106 (2017).

[12] N. Rivera and I. Kaminer, Nat. Rev. Phys. **2**, 538 (2020)

[13] A. Poddubny, I. Iorsh, P. Belov, and Y. Kivshar, Nat. Photonics **7**, 948 (2013).

[14] V. P. Drachev, V. A. Podolskiy, and A. V. Kildishev, Opt. Express **21**, 15048 (2013).

[15] F. Liu, L. Xiao, Y. Ye, M. Wang, K. Cui, X. Feng, W. Zhang and Y. Huang, Nat. Photonics **11**, 289 (2017).

[16] S. Liu, P. Zhang, W. Liu, S. Gong, R. Zhong, Y. Zhang, and M. Hu, Phys. Rev. Lett. **109**, 153902 (2021).

[17] S. Liu, C. Zhang, M. Hu, X. Chen, P. Zhang, S. Gong, T. Zhao, and R. Zhong, Appl. Phys. Lett. **104**, 201104 (2014).

[18] C. Yu and S. Liu, Phys. Rev. Appl. **12**, 054018 (2019).

[19] I. Kaminer, Y. T. Katan, H. Buljan, Y. Shen, O. Ilic, J. J. Lopez, L. J. Wong, J. D. Joannopoulos, and M. Soljacic, Nat. Commun. **7**, 11880 (2016).

[20] M. Dyakonov and M. Shur, Phys. Rev. Lett. **71**, 15 (1993).

[21] T. Otsuji, Y. M. Meziani, T. Nishimura, T. Suemitsu, W. Knap, E. Sano, T. Asano, and V. V. Popov, J. Phys.: Condens. Matter **20**, 38 (2008).

[22] F. Veksler, A. P. Teppe, V. Dmitriev, K. Yu., W. Knap, and M. S. Shur, Phys. Rev. B **73**, 125328 (2006).

[23] M. Dyakonov and M. Shur, IEEE Trans. Electron Dev. **43**, 3 (1996).

[24] T. A. Morgado and M. G. Silveirinha, Phys. Rev. Lett. **119**, 133901 (2017).

[25] D. Correas-Serrano and J. S. Gomez-Diaz, Phys. Rev. B **100**, 081410(R) (2019).

[26] Y. Dong, L. Xiong, I. Y. Phinney, Z. Sun, R. Jing, A. S. McLeod, S. Zhang, S. Liu, F. L. Ruta, H. Gao, Z. Dong, R. Pan, J. H. Edgar, P. Jarillo-Herrero, L. S. Levitov, A. J. Millis, M. M. Fogler, D. A. Bandurin & D. N. Basov, Nature **594**, 513 (2021).

[27] W. Zhao, S. Zhao, H. Li, S. Wang, S. Wang, M. Iqbal, S. Kahn, Y. Jiang, X. Xiao, S. Yoo, K. Watanabe, T. Aniguchi, A. Zettl & F. Wang, Nature, **594**, 517 (2021).

[28] B. E. A. Saleh and M. C. Teich, *Fundamentals of Photonics*, 2nd Edition (Wiley-Interscience, 2007).

[29] A. H. Castro Neto, F. Guinea, N. M. R. Peres, K. S. Novoselov, and A. K. Geim, Rev. Mod. Phys. **81**, 10 (2009).

[30] M. A. K. Othman, C. Guclu, and F. Capolino, Opt. Express **21**, 7614 (2013).

[31] I. V. Iorsh, I. S. Mukhin, I. V. Shadrivov, P. A. Belov, and Y. S. Kivshar, Phys. Rev. B **88**, 039904 (2013).

[32] J. Brouillte, G. Papadakis, and H. Atwater, Opt. Express **27**, 21 (2019).

[33] Y. C. Chang, C. H. Liu, C. H. Liu, S. Zhang, S. Marder, E. Narimanov, Z. Zhong, and T. B. Norris, Nat. Commun. **7**, 10568 (2016).

[34] J. S. Gómez-Díaz, M. Tymchenko, and A. Alù, Phys. Rev. Lett. **114**, 233901 (2015).

[35] J. S. Gomez-Diaz and A. Alù, ACS Photonics **3**, 12 (2016).

[36] E. Yoxall, M. Schnell, A. Y. Nikitin, O. Txoperena, A. Woessner, M. B. Lundeberg, F. Casanova, L. E. Hueso, F. H.L. Koppens, R. Hillenbrand, Nat. Photonics **9**, 675 (2015).

[37] P. Li, I. Dolado, F. Javier Alfaro-Mozaz, F. Casanova, L. E. Hueso, S. Liu, J. H. Edgar, A. Y. Nikitin, S. Vélez, and R. Hillenbrand, Science **359**, 892 (2018).

[38] Shankar R., *Principles of quantum mechanics, 1st* Edition (Springer Science & Business Media, 2012).

[39] D. Correas-Serrano, A. Alù, and J. S. Gómez-Díaz, Phys. Rev. B **96**, 075436 (2017).

[40] See Supplemental Information for a detailed description of the proposed formalism, including the development of the radiation rate formula shown in Eq. (1), as well as additional studies regarding the cut-off frequency of the emission and the influence of parameters as graphene's relaxation time and chemical potential, applied bias, geometrical dimensions, and electron velocity on the emission process.

[41] M. A. Yamoah, W. Yang, and E. Pop, ACS Nano **11**, 9914 (2017).

[42] D. Correas-Serrano, J. S. Gómez-Díaz, M. Tymchenko, and A. Alù, Opt. Express **23**, 29434 (2015).

[43] K. Y. Bliokh, F. Rodríguez-Fortuño, A. Y. Bekshaev, Y. S. Kivshar, and Franco Nori, Opt. Lett. **43**, 963 (2018).

[44] R. A. Maniyara, D. Rodrigo, R. Yu, J. Canet-Ferrer, D. S. Ghosh, R. Yongsunthon, D. E. Baker, A. Rezikyan, F. J. García de Abajo, and V. Pruneri, Nat. Photonics **13**, 328 (2019).

[45] S. Manzeli, D. Ovchinnikov, D. Pasquier, O. V. Yazyev, and A. Kis, Nat. Rev. Mat. **2**, 17033 (2017).

[46] D. Akinwande, C. Huyghebaert, C. H. Wang, M. I. Serna, S. Goossens, L. J. Li, H.S. Philip Wong, and F. H. L. Koppens, Nature **573**, 507 (2019).